\documentclass[twocolumn,a4paper,10pt,prl]{revtex4}%
\usepackage{amsmath}
\usepackage{graphicx}
\usepackage{amsfonts}
\usepackage{amssymb}%
\begin{document}
\title{Electromagnetically induced transparency and dynamic Stark effect in coupled
quantum resonators}
\author{M. A. de Ponte, C. J. Villas-B\^{o}as, R. M. Serra, and M. H. Y. Moussa}
\affiliation{Departamento de F\'{\i}sica, Universidade Federal de S\~{a}o Carlos, P.O. Box
676, S\~{a}o Carlos\textit{, 13565-905, S\~{a}o Paulo, }Brazil}

\begin{abstract}
In this work we reproduce the phenomenology of the electromagnetically induced
transparency and dynamic Stark effect in a dissipative system composed by two
coupled bosonic fields under linear and nonlinear amplification process. Such
a system can be used as a quantum switch in networks of oscillators.

PACS number: 32.80.-t, 42.50.Ct, 42.50.Dv

\end{abstract}
\maketitle

Much of the recent work in quantum optics has been devoted to quantum
coherence and interference as a result of the mastery gained in manipulating
atom-field interaction, which allows us to probe fundamental quantum effects.
Within this scenario, electromagnetically-induced transparency (EIT)
\cite{Harris} plays an essential role in a variety of processes, ranging from
lasing without inversion \cite{Zibrov} and enhanced nonlinear optics
\cite{Imamoglu} to quantum computation and communication \cite{Lukin}. Relying
on destructive quantum interference, EIT is a phenomenon in which\ the
absorption of a probe-laser field resonant with an atomic transition is
reduced or even eliminated by the application of a strong driving laser to an
adjacent transition. This effect can be understood as arising from the
ac-Stark splitting of an excited atomic state linking the two adjacent
transitions. When the splitting is smaller than the excited state width, the
two resulting levels are indistinguishable, leading to a destructive
interference in the probe absorption spectrum \cite{Harris}. As the intensity
of the driving field increases, the splitting is enlarged, eliminating the
indistinguishability and, consequently, the absorption spectrum evolves to the
Autler-Townes (AT) doublet. By a similar mechanism, although relying on
nonlinear interactions, the strong excitation of a two-level atom induces a
dynamic Stark effect (DSE) leading to the appearance of sidebands in the
resonance fluorescence spectrum \cite{Mollow}.

Recently, some effects of three-level atom optics, including EIT, were
simulated \cite{Eckert} \emph{via} the tunneling interaction in the context of
trapped atoms.\textbf{ }In this connection,\textbf{ }the phenomenology
observed in EIT was reproduced in a system of two coupled dissipative
classical oscillators in Ref. \cite{Paulo}. One of these oscillators, modeling
the three-level atom, is subject to a linear driving force \cite{Paulo}
playing the role of the probe field. The pump field is simulated by the
coupling between the oscillators. The quantum version of this classical analog
of EIT, which helps to deepen our understanding of this phenomenon and its
properties, will be analyzed in the present work. Here, the classical
oscillators are replaced by two quantum resonators interacting through a
Josephson-type coupling. Linear and nonlinear amplifications are employed as
the probe fields, the former is used to emulated the EIT phenomenon, and the
latter is applied to obtain the DSE --- reproducing the phenomenology of the
nonlinear mechanism related to resonance fluorescence from a strongly-driven
two-level transition. The coupling parameter between the resonators plays the
role of the driving field, by analogy with the standard EIT and DSE. We show
that the signatures of these processes can be obtained by measuring the field quadratures.

The Hamiltonian of the coupled dissipative resonators, labeled by $\ell=1,2$,
is given by ($\hbar=1$)
\begin{align}
H  &  =\sum_{\ell}\omega_{\ell}a_{\ell}^{\dagger}a_{\ell}+\sum_{\ell,k}%
\omega_{\ell k}b_{\ell k}^{\dagger}b_{\ell k}\nonumber\\
&  +\lambda\left(  a_{1}a_{2}^{\dagger}+a_{1}^{\dagger}a_{2}\right)
+\sum_{\ell,k}V_{\ell k}\left(  a_{\ell}b_{\ell k}^{\dagger}+a_{\ell}%
^{\dagger}b_{\ell k}\right) \nonumber\\
&  +\digamma\left[  \left(  \operatorname*{e}\nolimits^{-i\nu t}a_{1}%
^{\dagger}\right)  ^{\mathfrak{p}}+\left(  \operatorname*{e}{}^{i\nu t}%
a_{1}\right)  ^{\mathfrak{p}}\right]  , \label{Eq1}%
\end{align}
where $a_{\ell}^{\dagger}$ ($a_{\ell}$) is the creation (annihilation)
operator for the oscillator mode $\ell$ of frequency $\omega_{\ell}$, whereas
$b_{\ell k}^{\dagger}$ ($b_{\ell k}$) is the analogous operator for the $k$th
bath mode of oscillator $\ell$, whose corresponding frequency and coupling
strength are $\omega_{\ell k}$ and $V_{\ell k}$, respectively. The coupling
strength between the oscillators is $\lambda$ and the classical driving field
applied to oscillator $1$ has intensity $\digamma$ and frequency
$\mathfrak{p}\nu$, with the parameter $\mathfrak{p}=1$ or $2$ for the linear
or nonlinear amplification process.

Following the reasoning in Ref. \cite{Mickel}, we obtain (in a rotating frame
with frequency $\nu$ and assuming the reservoir at zero temperature) the
master equation for the system dynamics, given by
\begin{align}
\frac{d\rho_{12}\left(  t\right)  }{dt}  &  =i\left[  \rho_{12}\left(
t\right)  ,H_{0}\right] \nonumber\\
&  +\sum_{\ell}\frac{\Gamma_{\ell}}{2}\left(  \left[  a_{\ell}\rho_{12}\left(
t\right)  ,a_{\ell}^{\dagger}\right]  +\left[  a_{\ell},\rho_{12}\left(
t\right)  a_{\ell}^{\dagger}\right]  \right)  \mathrm{{,}} \label{Eq2}%
\end{align}
where $\Gamma_{\ell}$ is the damping rate of resonator $\ell$ and we have
defined
\begin{align}
H_{0}  &  \equiv\sum_{\ell}\Omega_{\ell}a_{\ell}^{\dagger}a_{\ell}%
+\lambda\left(  a_{1}a_{2}^{\dagger}+a_{1}^{\dagger}a_{2}\right) \nonumber\\
&  +\digamma\left[  \left(  a_{1}^{\dagger}\right)  ^{\mathfrak{p}}+\left(
a_{1}\right)  ^{\mathfrak{p}}\right]  \mathrm{{,}} \label{Eq3}%
\end{align}
with $\Omega_{\ell}\equiv\omega_{\ell}-\nu$ being the effective\ frequency\ of
the oscillator $\ell$ shifted by the amplification process. Using the standard
procedures and assuming Markovian white noise, we derive a $c$-number version
of the master equation (\ref{Eq2}) for the symmetric ordered characteristic
function $\chi(\left\{  \eta_{\ell}\right\}  ,t)$, given by%
\begin{align}
\frac{d\chi(\left\{  \eta_{\ell}\right\}  ,t)}{dt}  &  =\sum_{\ell}\left[
\left(  C_{\ell}(\left\{  \eta_{\ell}\right\}  )\frac{\partial}{\partial
\eta_{\ell}}+c.c.\right)  \right. \nonumber\\
&  \left.  +i\digamma\left(  \eta_{1}+\eta_{1}^{\ast}\right)  \left(  \eta
_{1}-\eta_{1}^{\ast}\right)  ^{\left(  \mathfrak{p}-1\right)  }\delta_{\ell
,1}\right] \nonumber\\
&  \times\chi(\left\{  \eta_{\ell}\right\}  ,t), \label{Eq5}%
\end{align}
where $C_{\ell}(\left\{  \eta_{\ell}\right\}  )=\eta_{\ell}\Lambda_{\ell
}+i\lambda\eta_{\ell-(-1)^{\ell}}+2i\left(  \mathfrak{p}-1\right)
\digamma\eta_{1}^{\ast}$, $\Lambda_{\ell}=\Gamma_{\ell}/2+i\Omega_{\ell}$, and
$\eta_{\ell}$ is the complex parameter of the characteristic function. From
the analytical solution of Eq. (\ref{Eq5}), obtained by standard techniques
\cite{Mickel}, we observe for the linear amplification ($\mathfrak{p}=1$),
that the modes in both resonators always evolve to stationary coherent states,
characterized by the equilibrium between the amplification process and the
dissipative mechanisms, mediated by the coupling strength $\lambda$. These
stationary coherent states, $\left\vert \alpha\right\rangle _{1}$ and
$\left\vert \beta\right\rangle _{2}$, have the complex amplitudes
$\alpha=-i\digamma\Lambda_{2}/\left(  \lambda^{2}+\Lambda_{1}\Lambda
_{2}\right)  $ and $\beta=-\lambda\digamma/\left(  \lambda^{2}+\Lambda
_{1}\Lambda_{2}\right)  $.

The EIT-like phenomenon, which occurs in the region where $\Gamma_{1}%
\gg\lambda\gg\Gamma_{2}$, is characterized by the state of the field in
resonator $1$ around the relaxation time $\tau_{R}$ of the joint system. For
the linear amplification case, the relaxation time, derived from the condition
$\partial\chi(\left\{  \eta_{\ell}\right\}  ,t)/\partial t\rightarrow0$
\cite{Mickel},\ is given by the simple expression $\tau_{R}=2\lambda
^{2}/\Gamma_{1}+\Gamma_{2}/2$. In Ref. \cite{Mickel}, it is noted that the
cavity-field states are interchanged between the resonators through state-swap
and the state-recurrence dynamics. Before the relaxation time, the coherence
dynamics of quantum states between the two resonators prevents the EIT from
occurring. Therefore, the system must achieve a stationary state to enable the
appearance of the destructive interference, which promotes the EIT. For this
reason, from here on we will be interested in the states of resonator $1$
around the relaxation time.

In atomic samples the EIT phenomenon is characterized by the permittivity of
the medium, associated with the complex quantity $n+ik$, where the dispersive
and the absorptive responses follow, respectively, from the real part of the
refractive index $n$ and the absorption coefficient $k$ \cite{Atomic}. Here,
the role of the permittivity is played by the quantity $\mathcal{E}_{\ell
}^{1/2}\operatorname*{e}\nolimits^{i\Theta_{\ell}}$ representing the center of
the quasi-probability distribution of the cavity-field state in the phase
space defined by its quadratures $X_{\ell}=${\Large \ }$\left(  a_{\ell
}+a_{\ell}^{\dagger}\right)  /2$ and{\Large \ }$Y_{\ell}=\left(  a_{\ell
}-a_{\ell}^{\dagger}\right)  /2i$. The absorption coefficient is directly
associated with the mean energy of the cavity field $\ell$, given by the
correlation function $\mathcal{E}_{\ell}=\left\langle X_{\ell}^{2}+Y_{\ell
}^{2}\right\rangle $ (in units of $\hbar\omega_{\ell}$). The dispersive
response is associated with the angle $\Theta$ through the correlation
function $\left\langle X_{\ell}Y_{\ell}+Y_{\ell}X_{\ell}\right\rangle
/\mathcal{E}_{\ell}=\sin\left(  2\Theta_{\ell}\right)  $.

In Ref. \cite{Paulo} the dispersive response $\mathrm{Re}\left[
x_{1}(\widetilde{\omega}-\widetilde{\omega}_{1})\right]  $ is derived from the
real part of the frequency dependence of the amplitude $x_{1}(\widetilde
{\omega}-\widetilde{\omega}_{1})$ of one oscillator (frequency $\widetilde
{\omega}_{1}$), submitted to a linear driving force (frequency $\widetilde
{\omega}$), and coupled to another oscillator (frequency $\widetilde{\omega
}_{2}$ and amplitude $x_{2}$). In the present model, the equivalent mean value
$\left\langle Y_{\ell}\right\rangle $ also leads to the required dispersive
response, as in \cite{Paulo}, for the linear amplification process
($\mathfrak{p}=1$). However, in the case of nonlinear amplification
($\mathfrak{p}=2$), the mean value $\left\langle Y_{\ell}\right\rangle $ is
not a convenient quantity to measure the dispersive response since it is null
when the initial excitation of the cavity field $\ell$ is also null (vacuum
state). To circumvent this problem, we have considered the correlation
function $\left\langle X_{\ell}Y_{\ell}+Y_{\ell}X_{\ell}\right\rangle
/\mathcal{E}_{\ell}$ to describe the dispersive response in a quite general
way. In fact, its classical analog, proportional to $\mathrm{Re}\left[
x_{\ell}p_{\ell}\right]  $ (where $p_{\ell}$ is the momentum of the classical
oscillator $\ell$), also leads to the usual dispersion exhibiting a very steep
slope in the EIT regime. The absorptive and dispersive responses given by the
correlation functions $\left\langle X_{\ell}^{2}+Y_{\ell}^{2}\right\rangle $
and $\left\langle X_{\ell}Y_{\ell}+Y_{\ell}X_{\ell}\right\rangle
/\mathcal{E}_{\ell}$, respectively, are computed from the characteristic
function $\chi(\left\{  \eta_{m}\right\}  ,t)$ ($m=1,2$) as
\begin{subequations}
\begin{align}
\mathcal{E}_{\ell}  &  =-\left.  \frac{\partial^{2}}{\partial\eta_{\ell
}\partial\eta_{\ell}^{\ast}}\chi(\left\{  \eta_{m}\right\}  ,t)\right\vert
_{\eta_{m}\rightarrow0}\mathrm{{,}}\label{Eq10a}\\
\sin\left(  2\Theta_{\ell}\right)   &  =\frac{1}{\mathcal{E}_{\ell}}\left.
\left(  \frac{\partial^{2}}{\partial\left(  \eta_{\ell}\right)  ^{2}}%
-\frac{\partial^{2}}{\partial\left(  \eta_{\ell}^{\ast}\right)  ^{2}}\right)
\chi(\left\{  \eta_{m}\right\}  ,t)\right\vert _{\eta_{m}\rightarrow
0}\mathrm{{.}} \label{Eq10b}%
\end{align}

\textit{Linear amplification}. To characterize the EIT, we plot the mean
values $\mathcal{E}_{1}$ and $\sin\left(  2\Theta_{1}\right)  $, for resonator
$1$, against the detuning $\Delta=\left(  \nu-\omega_{1}\right)  /\Gamma_{1}$.
Before the relaxation time ($\tau_{R}$), i.e., $t\ll\tau_{R}$, the EIT is
inhibited by the coherence dynamics between the cavity-field states. Near the
relaxation time, the EIT starts appearing, evolving to its standard shape at
the relaxation time, as displayed in Fig. $1($a$)$. In Fig. 1(b) we assume the
non-degenerate case $\omega_{1}=\omega_{2}-\epsilon$ at the relaxation time,
where the hole burned when $\nu=\omega_{1}$ in the degenerate case [Fig.
1(a)], shifts to $\nu=\omega_{1}+\epsilon$ (where $\epsilon/\Gamma_{1}=3$). It
is interesting to note, that in the non-degenerate case [Fig. 1(b)], the EIT
happens when the linear amplification ($\nu$) is in resonance with resonator 2
($\omega_{2}$). A complete transparency occurs only when $\Gamma_{2}=0$, since
the minimum of the hole burned in the peak of the absorptive response is given
by $\mathcal{E}_{1}^{\min}=\left[  2\digamma\Gamma_{2}/\left(  4\lambda
^{2}+\Gamma_{1}\Gamma_{2}\right)  \right]  ^{2}$. Assuming the same parameters
as those in Fig. $1$(a) but a stronger coupling strength ($\lambda/\Gamma
_{1}=1$), the EIT evolves to the AT regime as shown in Fig. 1(c). Finally, the
dispersion curves, in the right side of Fig. 1(a,b,c), display the well-known
signature of the EIT effect.%

\begin{figure}
[ptb]
\begin{center}
\includegraphics[
trim=0.200547in 3.167685in 0.167516in 0.489603in,
height=9.2346cm,
width=9.0281cm
]%
{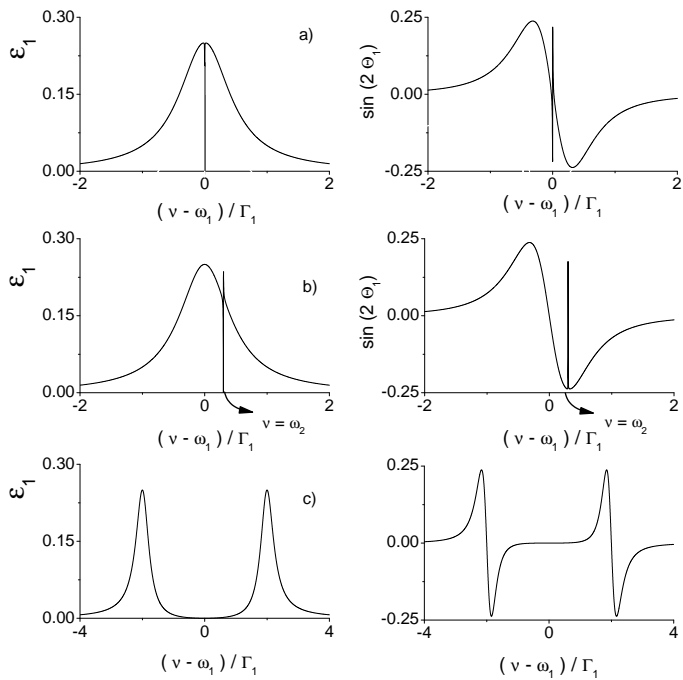}%
\caption{Absorptive-like (left side) and dispersive-like (right side) curves
for the linear amplification process (at $t=\tau_{R}$) assuming: (a) the
degenerate case ($\omega_{1}=\omega_{2}$) and $\Gamma_{2}$ $\ll\lambda\ll$
$\Gamma_{1}$, (b) the non-degenerate case ($\omega_{1}=\omega_{2}-\epsilon$),.
(c) the degenerate case and $\Gamma_{2}$ $\ll\lambda\approx$ $\Gamma_{1}$. To
plot these figures we have assumed the following parameters: $\Gamma
_{2}/\Gamma_{1}=10^{-4}$\textbf{ }and\textbf{ }$\digamma/\Gamma_{1}%
=1/4$.$\ \ $\textbf{ }For (a) and (b) we have used\textbf{ }$\lambda
/\Gamma_{1}=2\times10^{-2}$ and for (c) $\lambda/\Gamma_{1}=1$.}%
\end{center}
\end{figure}

\textit{Nonlinear amplification}. In the nonlinear or parametric amplification
case ($\mathfrak{p}=2$), there is a well-known threshold in the dynamics of
the system \cite{Walls,Salomon} which is reached when the rescaled
amplification parameter, $\xi=4\digamma/\Gamma_{1}$, equals the critical value
$\xi_{c}$ (which depends on the critical amplification amplitude $\digamma
_{c}$). We obtain three different dynamics for the solutions of Eq.
(\ref{Eq5}), which depend on whether the rescaled amplification is strong
($\xi>\xi_{c}$), critical ($\xi=\xi_{c}$), or weak ($\xi<\xi_{c}$). When
$\xi\geq\xi_{c}$ the asymptotic solution diverges for any physical quantity
whereas we obtain stationary solutions when $\xi<\xi_{c}$. In the degenerate
resonance case, $\omega_{1}=\omega_{2}=\omega$, the critical amplification
parameter is given either by $\xi_{c}=1+4\lambda^{2}/\left(  \Gamma_{1}%
\Gamma_{2}\right)  $ (if $\lambda<\Gamma_{2}/2$) or $\xi_{c}=1+\Gamma
_{2}/\Gamma_{1}$(if $\lambda\geq\Gamma_{2}/2$). Evidently, when $\lambda=0$ we
recover the usual result $\xi_{c}=1$ for an uncoupled system under nonlinear
amplification. In these three regimes the state of oscillator $1$ evolves to a
squeezed coherent state (SCS) which is stationary for $\xi<\xi_{c}$ and whose
energies increase monotonically for $\xi\geq\xi_{c}$.

Now, in order to analyze the behavior of the absorptive and dispersive curves
of resonator $1$, in the three regimes exposed above, we plot the rate
$\mathcal{E}_{1}/\mathcal{E}$, instead of $\mathcal{E}_{1}$, where
$\mathcal{E}$ stands for the value of $\mathcal{E}_{1}$ when the detuning
$\Delta=0$. In this case, the EIT occurs only in a narrow interval $\left\vert
\xi\right\vert \lesssim\xi_{c}+10^{-4}$ around the threshold given by the
critical parameter $\xi_{c}$, i.e., around $\digamma_{c}=(\Gamma_{1}%
+4\lambda^{2}/\Gamma_{2})/4$ for $\lambda<\Gamma_{2}/2$, or $\digamma
_{c}=(\Gamma_{1}+\Gamma_{2})/4$ for $\lambda\geq\Gamma_{2}/2$. Differently
from the linear amplification process, where the EIT can be achieved
independently of the ratio $\digamma/\Gamma_{1}$, the strong dependence of
this effect upon the amplitude of the nonlinear driving field follows from the
two-photon nature of this amplification process. As far as the coupling
strength is concerned, although it must be much smaller than $\Gamma_{1}$ as
in the linear amplification case, here it can be around $\Gamma_{2}$, obeying
the relation $\Gamma_{2}$ $\lesssim\lambda\ll$ $\Gamma_{1}$.%

\begin{figure}
[ptb]
\begin{center}
\includegraphics[
trim=0.254027in 5.270031in 0.307506in 0.878338in,
height=5.9968cm,
width=8.4175cm
]%
{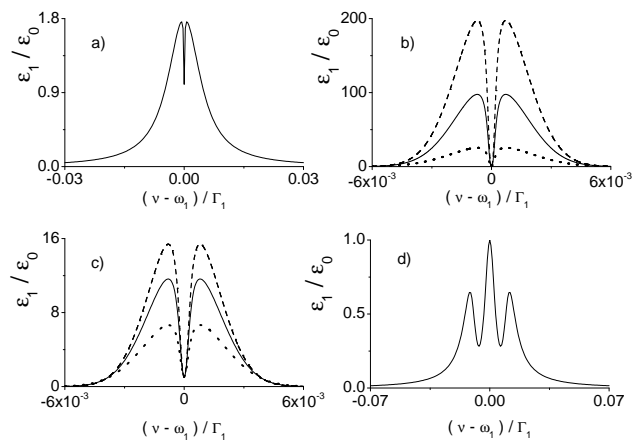}%
\caption{Absorptive-like curves for the nonlinear amplification process
($\omega_{1}=\omega_{2}$): in (a) the weak regime ($\xi=\xi_{c}-1.5\times
10^{-4},$ $\xi_{c}=1+10^{-4}$ ), (b) critical regime ($\xi=\xi_{c}$), (c)
strong regime ($\xi=\xi_{c}+1,5\times10^{-4}$), and (d) in the weak
amplification regime at the relaxation time $\widetilde{\tau}_{R}$, but
increasing the coupling strength. To plot these figures we have assumed the
following parameter: $\Gamma_{2}/\Gamma_{1}=10^{-4}$.$\ \ $\textbf{ }For (a-c)
we have used\textbf{ }$\lambda/\Gamma_{1}=5\times10^{-5}$ and for (d) we have
used $\lambda/\Gamma_{1}=10^{-2}$. Where the dotted line is for $t=0.8\times
\widetilde{\tau}_{R}$, the solid line is for $t=\widetilde{\tau}_{R}$, the
dashed line is for $t=1.1\times\widetilde{\tau}_{R}$, and we have assumed the
vacuum states as initial states for the resonators 1 and 2. }%
\end{center}
\end{figure}

In the nonlinear case the relaxation time $\widetilde{\tau}_{R}$ can only be
obtained\textbf{ }from the condition $\partial\chi(\left\{  \eta_{\ell
}\right\}  ,t)/\partial t\rightarrow0$ in the weak amplification regime. This
time will be used as a reference to compare the three different amplification
regimes, since there is no stationary dynamics when $\xi\geq\xi_{c}$. We can
see in Fig. $2($a$)$, the absorptive curve for the weak amplification regime,
that the EIT is not so pronounced as in the linear amplification case. In
fact, it can barely be characterized as an EIT, since the maximum of the ratio
$\mathcal{E}_{1}/\mathcal{E}_{0}$ is close to unity (i.e., the minimum of the
hole burned in the peak of the absorptive curve, when $\Delta=0$). The
absorptive curve remains immovable for $t>\widetilde{\tau}_{R}$ due to the
stationary solution achieved in the weak amplification regime. Differently
from Fig. $2($a$)$, we observe that in the critical regime, Fig. $2$(b), the
maximum of the ratio $\mathcal{E}_{1}/\mathcal{E}$ is about two orders of
magnitude larger than unity ($\mathcal{E}_{1}=\mathcal{E}$) for $t=\widetilde
{\tau}_{R}$ (solid line), increasing as time goes on. However, as in the weak
amplification regime, the behavior observed here can hardly be characterized
as an EIT, but rather a mixing of EIT and AT effects, since the hole burned in
the peak of the absorptive response presents a significant width compared to
the usual EIT. In Fig. $2($c$)$, for the strong amplification regime, a mixing
of the EIT and the AT effects is observed again, but the ratio $\mathcal{E}%
_{1}/\mathcal{E}$ does not increase as quickly as in the critical regime ---
even so, $\mathcal{E}_{1}$ and $\mathcal{E}$ increase, independently, as
faster as they do in the critical regime. The maximum of the rate
$\mathcal{E}_{1}/\mathcal{E}$ is (only) about one order of magnitude larger
than unity ($\mathcal{E}_{1}=\mathcal{E}$) for $t=\widetilde{\tau}_{R}$ and
increases as the time goes on. We note that the range of the detuning $\Delta
$, in Figs. $2($b$)$ and $($c$)$ ---where the absorptive curves assume
significant values--- is considerably smaller than that in Fig. $2($a$)$ for
the weak regime. Therefore, as the rescaled amplification parameter $\xi$
increases, the range $\Delta$ of the absorptive curve decreases. From this
perspective, it is not difficult to understand why the range $\Delta$ of the
absorptive curve in the linear amplification regime (Fig. 1) is larger than
those in the nonlinear one. Regarding the dispersive curve, it happens to be
practically the same for the three parametric amplification regimes and the
steep slope characteristic of the EIT effect is lost.

Finally, in Fig. $2$(d) we display the absorptive curve in the weak
amplification regime at the relaxation time $\widetilde{\tau}_{R}$, but
increasing the coupling strength such that $\lambda/\Gamma_{1}=10^{-2}$
(instead of $5\times10^{-5}$). We thus observe the well-known pattern
exhibited by the resonance fluorescence spectrum of a two-level atom driven by
an incident field whose Rabi frequency is comparable to, or larger than, the
atomic linewidth. After all, as the phenomenology of the EIT can be reproduced
in a system of two coupled dissipative oscillators, the same occurs with the
phenomenology of resonance fluorescence. Resonator $1$ again plays the role of
the two-level atom, while the strong driving field (whose intensity leads to a
modulation of the quantum dipole moment inducing sidebands in the spectrum
\cite{Scully}) is simulated by the nonlinear amplification together with
resonator $2$. In fact, high-intensity driving fields give rise to nonlinear
interactions which may be attributed to two laser fields \cite{Walls1},
modeled by the actual parametric field plus resonator $2$. The pattern
displayed in Fig. $2($d$)$ can also be obtained in the critical and strong
amplification regimes but, of course, it does not occur with linear
amplification, whatever its intensity. The parametric amplification field is
the only nonlinear ingredient in our system of two coupled dissipative oscillators.

We have shown that a system of two coupled dissipative oscillators, under
linear and parametric amplification, can reproduce the phenomenology of
EIT\ and DSE, respectively. Such phenomena can be, at principle, implemented
in several physical contexts, such as: cavity quantum electrodynamics
\cite{QED}, trapped ions \cite{Ions}, and nanomechanical oscillators
\cite{Nano}. The system presented here can be used for quantum switching
purposes, since the hole burned in the EIT regime can be shifted by external
means, feeding or draining the field in resonator 1. For example, the
dispersive interaction of a fermionic particle with the field in oscillator 2
can pull both oscillators out of resonance, changing the profile of the
absorption curve of oscillator 1 from Fig. 1(a) to 1(b). Therefore, these
effects have the potential to be used as a control mechanism in networks of
quantum oscillators, a subject which attracts great attention nowadays
\cite{Plenio}.
\end{subequations}
\begin{acknowledgments}
This work was supported by FAPESP and CNPq, Brazilian agencies. We also thank
Profs. P. Nussenzveig and B. Baseia for helpful discussions.
\end{acknowledgments}

\end{document}